# Spin disorder in a stacking polytype of a layered magnet

Xianghan Xu[1*], Guangming Cheng[2], Danrui Ni[1], Xin Gui[3], Weiwei Xie[4], Nan Yao[2], and R. J. Cava[1*]

1. Department of Chemistry, Princeton University, Princeton, NJ 08544
2. Princeton Institute for the Science and Technology of Materials, Princeton University, Princeton, NJ 08544
3. Department of Chemistry, University of Pittsburgh, Pittsburgh, PA 15260
4. Department of Chemistry, Michigan State University, Lansing, MI 48824

**Abstract**

Strongly correlated ground states and exotic quasiparticle excitations in low-dimensional systems are central research topics in the solid state research community. The present work develops a new layered material and explores the physical properties. Single crystals of 3R-$Na_2MnTeO_6$ were synthesized via a flux method. Single crystal x-ray diffraction and transmission electron microscopy reveal a crystal structure with ABC-type stacking and an *R*-3 space group, which establishes this material as a stacking polytype to previously reported 2H-$Na_2MnTeO_6$. Magnetic and heat capacity measurements demonstrate dominant antiferromagnetic interactions, the absence of long-range magnetic order down to 0.5 K, and field-dependent short range magnetic correlations. A structural transition at ~ 23 K observed in dielectric measurements may be related to displacements of the Na positions. Our results demonstrate that 3R-$Na_2MnTeO_6$ displays low-dimensional magnetism, disordered structure and spins, and the system displays a rich structure variety.

## I. Introduction

Low-dimensional magnetic systems have attracted research interest in recent years due to their potential for exhibiting spin entanglement and magnetic quantum states. For 1D or 2D Heisenberg spins, the ground state can be prevented from attaining long-range ordering at finite temperature, as proposed by the Mermin-Wagner theorem [1]. Long-range ordering can also be suppressed for Ising spins, through geometric frustration, leading to an entangled ground state [2,3]. Nevertheless, in the bulk crystals of many layered frustrated magnets, disorder or non-negligible interlayer exchange interactions often introduce spin freezing or long-range ordering in spite of expectations for layered systems [4-8].

The crystallographic symmetry of layered materials strongly depends on the stacking type of the atom layers. In the closest packing case, 'AB' stacking results in a hexagonal cell (2H), while 'ABC' stacking produces a rhombohedral cell (3R) [9]. Other types of stacking also create polytypes, such as are seen for trigonal (1T) and monoclinic (1T') TaSe$_2$ [10]. Different stacking types can play an important role in the electronic properties of layered materials as well. For instance, the 2H phase of MoTe$_2$ is semiconducting while the 1T′ phase is semi-metallic [11]. In the Kitaev spin liquid candidate α-RuCl$_3$, stacking faults are believed to cause an anomalous magnetodielectric response [12]. Recently, ABC stacked rhombohedral tri-layer graphene has been found to exhibit superconductivity without being twisted [13], which is not the case for other stacking variants.

Despite the above-mentioned novel physics stemming from differences in stacking type, the interplay between stacking type and magnetism in layered magnetic insulators has not been well studied. This is partially due to the lack of magnetic layered oxide materials that display stacking polytypes. The $A_2MTeO_6$ (A = Li, Na and M = Ti, Mn, Sn, Ge) family contains stacked 2D honeycomb structure units [14,15]. $Na_2GeTeO_6$ exhibits both 2H and 3R polytypes, suggesting that this layered oxide family can display structural stacking variability. Recently, a magnetic compound, 2H-$Na_2MnTeO_6$, has been described, with a commensurate 120º helical order found below 5.5 K [16]. However, the synthesis of the 3R version of $Na_2MnTeO_6$ in single-phase form has been argued to be infeasible [17]. Therefore, the synthesis of magnetic 3R compounds in this family is important but not reported to date.

Here we experimentally demonstrate that the 'ABC' stacking type of a honeycomb layered magnet $Na_2MnTeO_6$ leads to a new $R$-3 lattice (3R) as well as magnetism that is distinct from the previously reported hexagonal 'AB' stacking (2H) compound with the same chemical formula. Single crystals of 3R-$Na_2MnTeO_6$ are grown using a flux method. Magnetic and heat capacity study reveals a highly correlated ground state but no long-range magnetic ordering. Single-crystal x-ray and electron diffraction studies find a disordered distribution of $Mn^{4+}$ and $Te^{6+}$ ions in the honeycomb layer of the $R$-3 lattice. Magnetic measurements rule out long-range ordering down to 1.8 K. Heat capacity measurements confirm the absence of long-range order down to 0.5 K and unveil the presence of field-dependent short-range correlations. Thus, we argue that 3R-$Na_2MnTeO_6$ provides a high-quality platform for studying the magnetic excitation continuum in low dimension and the interplay between inter-site disorder and spin disordered states.

## II. Experimental

Single crystal of 3R-$Na_2MnTeO_6$ was grown by a flux method. $Na_2CO_3$ (99.5%, Alfa Aesar), $Mn_2O_3$ (98%, Alfa Aesar), and $TeO_2$ (99.99%, Alfa Aesar) powders in molar ratio 6: 1: 12 were mixed, placed in a platinum crucible, heated to 850ºC in air, soaked at that temperature for 10 hours, cooled to 500ºC at 3ºC/h, and then cooled at 100ºC/h to room temperature. The 3R-$Na_2MnTeO_6$ crystals with hexagonal plate shape and typical size 3 mm * 3 mm * 50 μm were mechanically separated after bathing the product in a 1M NaOH water solution overnight. Similar flux growth method was used to grow 3R-$Li_2GeTeO_6$ crystals with a slower cooling rate [18].

Single crystal diffraction was performed in a Bruker single crystal X-ray diffractometer. The refinement was performed by using the SHELXTL Software Package [19,20]. Scanning electron microscopy (SEM) images were collected using a Quanta 200 FEG Environmental-SEM. Thin lamellae were prepared for TEM study by focused ion beam (FIB) cutting using a Helios NanoLab G3 UC dual-beam FIB/SEM system. Sample thinning was accomplished by gently polishing the sample using a 2 kV gallium ion beam in order to minimize surface damage caused by the high-energy focused ion beam. Conventional TEM imaging, SAED, atomic-resolution HAADF-

STEM imaging and energy dispersive X-ray spectroscopy (EDS) mapping were performed on a Titan Cubed Themis 300 double Cs-corrected STEM equipped with an extreme field emission gun source operated at 300 kV with a super-X EDS system. The system was operated at 300 kV.

The temperature-dependent magnetization and heat capacity measurements were performed in a Quantum Design PPMS-9 Dynacool. A Hellium-3 refrigerator was used to collect the heat capacity data below 2 K. The dielectric permittivity was measured using a QuadTech 7600 LCR meter externally connected to the PPMS probe. The electrodes were made of silver epoxy.

## III. Results and Discussion
### 3.1 Crystal structure

A photograph of as-grown $Na_2MnTeO_6$ crystal is displayed in Fig. 1a left inset. The crystals are highly cleavable. The SEM image (Fig. 1a right inset) displays an atomically flat cleaved surface and terraces, features that are commonly observed in Van der Waals materials. Thin flakes of these crystals are reddish and transparent in transmitted natural light, a characteristic of their optical band gap. As shown in Fig. 1a, the powder XRD pattern of ground crystals can be fit well by a rhombohedral unit cell with $a$ = 5.219(1) Å and $c$ = 15.91(5) Å (Fig. 1b), but not the hexagonal unit cell reported for 2H-$Na_2MnTeO_6$ (Fig. 1c). These results indicate that a new polytype has been formed. A tiny crystal flake was selected for single crystal x-ray diffraction structure determination. The refinement result confirms the $R$-3 symmetry with an ABC-type stacking of disordered Mn-Te layers, which makes the material isostructural to a reported $R$-3 polytype of $Na_2GeTeO_6$ [15]. Introducing Mn-Te ordering creates two distinct sites in the honeycomb layer, increases the in-plane cell dimension and lowers the lattice symmetry. Attempts to refine the crystal structure using such a model increased the $R_1$ value from 3.70% to 5.84%. If disorder is allowed in the model, the refinement gives occupancies Mn0.58Te0.42 in honeycomb site 1 and Mn0.42Te0.58 in honeycomb site 2. This result indicates that the Mn and Te in 3R-$Na_2MnTeO_6$ crystals are mostly statistically disordered (Fig. 1d), and that the lattice symmetry is closer to being centrosymmetric $R$-3. Therefore, a $R$-3 model with mixed Mn and Te in a ratio fixed to 0.5 : 0.5 was used to refine the SC-XRD data. The reported Mn-Te disorder level is around 15% in 2H-$Na_2MnTeO_6$ (Fig. 1e) [16], which is much smaller than observed in 3R-$Na_2MnTeO_6$ as reported here. The powder XRD on crushed crystals shows sharp (00L) peaks without splitting in addition to $K_{\alpha 1}$ and $K_{\alpha 2}$, which are consistent with the rhombohedral symmetry (Fig. S1) [21]. The crystallographic data for 3R-$Na_2MnTeO_6$, including atomic positions, site occupancies, and refined thermal parameters, are listed in Tables 1 and 2. The unit cell parameters obtained from single crystal diffraction match well with the ones obtained from the ground crystal powder XRD. The crystal structure is registered with the Cambridge Structural Database, number 2216668.

To further study the new stacking polytype (especially the local symmetry), we examined thin lamellar samples cut from single crystal 3R-$Na_2MnTeO_6$ by scanning/transmission electron microscopy (S/TEM), as shown in Fig. 2. A low-magnification TEM image of a studied sample area is displayed in Fig. 2a. The diffraction pattern in Fig. 2b verified that the polytype exhibits a *R*-3 structure which is consistent with the analysis from single crystal X-ray diffraction, with localized Mn-Te disordering. An atomic-resolution STEM image shown in Fig. 2c confirmed the 3R atomic stacking in the structure, viewed from [110] direction. Note that Na atoms cannot be imaged in the STEM image which is due to their sensitivity to high-energy electron beam. The neighboring atomic spots display the same image contrast which indicates a disordered distribution of Mn and Te. It should be mentioned that if Mn and Te atoms are orderly distributed in the structure, they will show clear image contrast because of the large difference in atomic number. In addition, stacking faults are observed in a few regions of the TEM lamellae. Fig. S2b displays the STEM atomic imaging of an area with stacking faults [21], and the corresponding diffraction pattern shows satellite spots between some of the rhombohedral 00l diffraction peaks. The Mn and Te atoms are still disorderly distributed in the stacking faults. Overall, both diffraction analysis and atomic-resolution imaging confirmed a disordered Mn-Te distribution in the single crystal, which is consistent with the single crystal x-ray diffraction refinement result.

**3.2 Magnetism**

The magnetic susceptibility measured along the *c* and *a* directions under an applied field of 1000 Oe is displayed in Fig. 3a. No long-range magnetic ordering is detected down to 1.8 K. The inset depicts the inverse susceptibility, for which a bending below 4 K along *a* reflects the onset of a short-range magnetic correlations. Curie-Weiss (CW) fitting gives $\theta_c = -3.0(4)$ K, $\mu_{eff,c} = 4.0(5)$ $\mu_B$, $\theta_{ab} = -4.1(7)$ K, and $\mu_{eff,ab} = 3.9(9)$ $\mu_B$/formula unit in the temperature range 7 K to 20 K, and $\theta_c = -4.9(5)$ K, $\mu_{eff,c} = 3.8(3)$ $\mu_B$/formula unit, $\theta_{ab} = -6.2(3)$ K, and $\mu_{eff,ab} = 3.9(5)$ $\mu_B$/formula unit in the temperature range 100 K to 150 K. The negative CW temperatures are an indication of dominant AFM interactions both interlayer and intralayer. The magnetic susceptibility of polycrystalline sample is calculated by $\chi_{poly} = \frac{1}{3}\chi_c + \frac{2}{3}\chi_{ab}$. The CW fitting to that average gives $\theta_{poly} = -3.73(0)$ K, $\mu_{eff,poly} = 4.01(0)$ $\mu_B$/formula unit in the temperature range 7 K – 20 K, and $\theta_{poly} = -9.2(3)$ K, $\mu_{eff,poly} = 3.8(5)$ $\mu_B$/formula unit in the temperature range 200 K – 300 K. The effective moments obtained from $\chi_{poly}$ are in good agreement with the expected spin only value of 3.87 $\mu_B$ for the $Mn^{4+}$ (S=3/2) free ion. (The inverse susceptibility as a function of temperature curves for H//c, H//a, and calculated polycrystalline sample are plotted in Fig. S3 [21].) Fig. 3b shows the isothermal magnetization versus field curve. Weak easy-axis anisotropy is observed, and no field-induced transitions are seen. In contrast, 2H-Na$_2$MnTeO$_6$ is reported to have a long-range ordering transition at around 5.5 K that can be unhidden in the temperature-dependent susceptibility measured at high applied fields [16]. Similar measurements were performed on a 3R-Na$_2$MnTeO$_6$ crystal along both the *a* and *c* directions (Fig. S4) [21], but the results do not show any sharp transitions in fields up to 9 T. As shown in Fig. S5a-S5b [21], the field cooling (FC) and zero field cooling (ZFC) susceptibility data measured at a field of 50 Oe don't show any detectable bifurcation, which rules out the possibilities of ferromagnetic moments or spin glassiness down to 1.8 K. The AC susceptibility was measured down to 1.8 K, and the results are shown in Fig. S5c-S5d [21]. The AC susceptibility $\chi'_{AC}$ doesn't show any anomaly or frequency dependence down to 1.8 K in zero DC field, while applying a 3 T field leads to a broad maximum at around 5 K in $\chi'_{AC}$, implying that the material displays field-enhanced short-range magnetic ordering.

### 3.3 Heat Capacity

To further confirm the absence of long-range ordering and look for underlying quantum states, heat capacity measurements were performed down to 0.5 K. Fig. 4a displays the heat capacity of 3R-Na$_2$MnTeO$_6$ up to 170 K in zero field. The phonon contribution dominates in the high temperature regime. However, it is found that the data cannot be well fitted by a Debye equation with one single Debye temperature. Therefore, a modified Debye equation, $C_{phonon} = 9R \sum_{n=1}^{2} C_n \left( \frac{T}{\Theta_{Dn}} \right)^3 \int_0^{\Theta_{Dn}/T} \frac{x^4 e^x}{(e^x - 1)^2} dx$, with two components is used to fit the heat capacity data. The constraint is that $C_1 + C_2 = 10$, which is the total number of atoms in the chemical formula Na$_2$MnTeO$_6$. The fitting curve is displayed in red, and gives fitting parameters $C_1 = 4.75$, $\Theta_{D1} = 1196.1$ K, and $C_2 = 5.25$, $\Theta_{D2} = 349.6$ K. Note that the same two-component Debye equation has been applied to fit the high-temperature heat capacity of other oxide solids containing heavy tellurium atoms [22,23]. Below 30 K, a broad peak with a maximum at around 5 K shows up and the heat capacity deviates from the phonon fitting, indicating the appearance of short-range magnetic correlations. After subtracting the phonon contribution, the magnetic entropy is obtained from integration and plotted in the inset of Fig. 4a. The magnetic entropy saturates at around 60 K and reaches 9.57 J·mole$^{-1}$·K$^{-1}$. This value is 83% of the theoretical total magnetic entropy 11.52 J·mole$^{-1}$·K$^{-1}$ for S=3/2 Mn$^{4+}$. The difference could be due to either an overestimated phonon contribution, or spin fluctuations at lower temperatures that lift the ground-state degeneracy [24]. As shown in Fig. 4b, applying magnetic fields gradually shifts the broad peak to higher temperatures. Consistently, Fig. 4c shows that the peak in $C_p/T$, i.e., the major contribution of the magnetic entropy, can be shifted to higher temperatures by applying fields. This behavior has been commonly observed in frustrated magnets [25,26]. Note that even at zero field, the broad peak in $C_p/T$ still exists; at a temperature of about 0.8 K. A comparison of the zero-field heat capacity of

3R-Na$_2$MnTeO$_6$ and previously published 2H-Na$_2$MnTeO$_6$ is displayed in Fig. S6 [21]. Obviously, 3R-Na$_2$MnTeO$_6$ shows a much broader maximum than long-range ordered 2H-Na$_2$MnTeO$_6$. We speculate that either different interlayer exchange interactions in the different stacking types or the Mn-Te disorder may account for the more disordered low temperature spin state in 3R-Na$_2$MnTeO$_6$ compared to that of 2H-Na$_2$MnTeO$_6$.

### 3.4 Discussion

Previous work reports the presence of a very small amount of 3R-Na$_2$MnTeO$_6$ for polycrystalline material synthesized at 600°C – 650°C; but that the 3R phase irreversibly transforms to the 2H phase after heating at 700°C – 750°C, indicating that the 3R phase is only stable at low temperature [17]. However, high temperature is critical for activating atom diffusion in the conventional solid state reaction method, which may explain why single-phase 3R-Na$_2$MnTeO$_6$ has not been synthesized previously. The single crystal growth method reported in this work can overcome this issue and a pure 3R phase is finally obtained. To do that, the raw materials were first melted at 850°C to ensure a uniform mixture, but the 3R phase did not immediately form. Instead, the phase forms crystals when the liquid is slowly cooled into the temperature range for which the 3R phase is thermodynamically most stable. Quenching the crucible at 650°C results in tiny 3R crystals, which indicates that 650°C is pretty close to the onset temperature of 3R crystallization. On the other hand, the relatively low crystallization temperature of 3R phase may count for the Mn-Te ionic disorder possibly because ions don't have enough activation energy to order. Similar tendency that the low-temperature phase tends to show more chemical disorder has been observed in the LiCoO$_2$ system. In the cubic polymorph synthesized at a low temperature, the Li and Co ions exhibit site disorder [27-30], but in the high-temperature rhombohedral polymorph, the Li and Co ions fully order [31]. The Mn-Te disorder results in a centrosymmetric $R$-3 space group crystal structure that is isostructural to the reported compound 3R-Na$_2$GeTeO$_6$ which is also fully Ge-Te disordered [15]. In contrast, the analogous compound Li$_2$GeTeO$_6$ displays a 93% ordered Ge-Te distribution within a rhombohedral space group [32]. In some previously reported materials with a high temperature order-disorder transition, nano domains of ordered regions can be observed in a TEM [33]. However, though the electron diffraction patterns unveil some local ordered regions for the current material, no domain features are observed in the real-space TEM images of 3R-Na$_2$MnTeO$_6$, which suggests that a high temperature order-disorder transition doesn't happen in this case.

Single-ion anisotropy and spin degeneracy play central roles in determining the low temperature ground states of frustrated magnets. Quantum fluctuations are typically most significant in systems with $S = 1/2$, leading to a highly entangled quantum spin liquid state [34]. As a special case, Co$^{2+}$ has spin 3/2, but the participation of spin-orbit coupling and crystal electric field can generate an effective doublet spin at low temperature[35-38]. In Cr$^{3+}$, Mn$^{4+}$, and Fe$^{5+}$ systems with larger spin values, geometrically frustrated antiferromagnetic interactions may also lead to spin disordered states. Examples include MgCr$_2$O$_4$ (S = 3/2 for Cr$^{3+}$) [39], BaTi$_{1/2}$Mn$_{1/2}$O$_3$ (S = 3/2 for Mn$^{4+}$) [40], and Li$_9$Fe$_3$(P$_2$O$_7$)$_3$(PO$_4$)$_2$ (S = 5/2 for Fe$^{3+}$) [41]. In 3R-Na$_2$MnTeO$_6$, the Mn ion has a formal valence of +4 and S = 3/2. The effective magnetic moment obtained from magnetic data in all temperature ranges is very close to that value (See Section 3.2), which suggests the orbital contribution to the effective moment is negligible.

As shown in Fig. 5, the dielectric permittivity of 3R-$Na_2MnTeO_6$ shows a subtle anomaly at ~ 23 K, marked by the dashed line. An applied magnetic field does not affect this feature at all, either out-of-plane or in-plane (similar, not shown), implying that this transition is not induced by the coupling of crystal structure and short-range magnetic correlations, but instead is purely structural. Note that the anomaly at around 250 K is likely from a dipole freezing and may not be intrinsic. In another sodium-containing layered tellurate $Na_2Ni_2TeO_6$, it was found that the intralayer diffusion of Na ions could create Na-ordering patterns and further lower the lattice symmetry [42], which provides a possible scenario for a structural transition in 3R-$Na_2MnTeO_6$. Note that no clear anomalies are observed in the magnetic susceptibility or heat capacity versus temperature data at 23 K, suggesting that the change of structure across this transition is subtle, such as slight displacements of the Na positions. In addition, though the room temperature space group *R*-3 is centrosymmetric, all the mirror symmetries are broken, and ferro-rotation order [43] is a possibility. The single crystals obtained here therefore provide excellent platforms for future studies concerning ferro-rotation and ferro-rotational domain configurations using optical techniques such as second harmonic generation [44].

**IV. Conclusion**

Crystals of 3R-$Na_2MnTeO_6$ were synthesized using a flux method and their properties are reported. Structure analysis by x-ray and electron diffraction reveals a *R*-3 space group and ABC- type stacking crystal structure, i.e., a new 3R stacking polytype compared to a reported 2H phase. Magnetic and heat capacity measurements are characteristic of low temperature short-range spin correlations and the absence of long-range spin order down to 0.5 K. Our study reflects the fact that the stacking of a layered oxide can be determined by varying the synthesis method. Our results also demonstrate a spin disordered state in a low-dimensional honeycomb lattice at low temperature. Considering the non-negligible level of Mn-Te chemical disorder, it cannot be concluded that whether the different magnetism of 2H- and 3R- $Na_2MnTeO_6$ is dominated by stacking types or chemical disorder. Therefore, future research should focus on how to diminish the chemical disorder by optimizing the flux growth condition such as a slower cooling rate or a careful annealing of the as-grown crystals. In addition, inspired by the fact that 3R-$Li_2GeTeO_6$ shows much better Ge-Te ordering than 3R-$Na_2GeTeO_6$, introducing Li in 3R-$Na_2MnTeO_6$, i.e., 3R-$Li_xNa_{2-x}MnTeO_6$, may help Mn and Te ions to order. The growth of 3R-$Na_2MnTeO_6$ single crystal brings up new opportunities such as searching for a magnetic excitation continuum, exotic local structures formed by interlayer ions, studies of structural ferroic order by advanced optical techniques and may initiate further exploration of the stacking polytypes of layered oxides.

**Acknowledgements**

This research was primarily supported by the Gordon and Betty Moore Foundation, grant GBMF-9066. The authors acknowledge the use of Princeton's Imaging and Analysis Center (IAC), which is partially supported by the Princeton Center for Complex Materials (PCCM), a National Science Foundation (NSF) Materials Research Science and Engineering Center (MRSEC; DMR-2011750). The single crystal diffraction data were obtained in a laboratory at Rutgers University, supported by the U.S. DOE-BES under Contract DE SC0022156.

**References**
[1] N. D. Mermin and H. Wagner, Absence of Ferromagnetism or Antiferromagnetism in One- or Two-Dimensional Isotropic Heisenberg Models, Phys. Rev. Lett. **17**, 1133 (1966).
[2] L. Balents, Spin liquids in frustrated magnets, Nature **464**, 199 (2010).
[3] C. Broholm, R. J. Cava, S. A. Kivelson, D. G. Nocera, M. R. Norman, and T. Senthil, Quantum spin liquids, Science **367**, eaay0668 (2020).
[4] S. H. Lee, H. Kikuchi, Y. Qiu, B. Lake, Q. Huang, K. Habicht, and K. Kiefer, Quantum-spin-liquid states in the two-dimensional kagome antiferromagnets $Zn_xCu_{4−x}(OD)_6Cl_2$, Nat. Mater. **6**, 853 (2007).
[5] Y. Shirata, H. Tanaka, A. Matsuo, and K. Kindo, Experimental Realization of a Spin-1/2 Triangular-Lattice Heisenberg Antiferromagnet, Phys. Rev. Lett. **108**, 057205 (2012).
[6] J. J. Wen, S. M. Koohpayeh, K. A. Ross, B. A. Trump, T. M. McQueen, K. Kimura, S. Nakatsuji, Y. Qiu, D. M. Pajerowski, J. R. D. Copley, and C. L. Broholm, Disordered Route to the Coulomb Quantum Spin Liquid: Random Transverse Fields on Spin Ice in $Pr_2Zr_2O_7$, Phys. Rev. Lett. **118**, 107206 (2017).
[7] T. Furukawa, K. Miyagawa, T. Itou, M. Ito, H. Taniguchi, M. Saito, S. Iguchi, T. Sasaki, and K. Kanoda, Quantum Spin Liquid Emerging from Antiferromagnetic Order by Introducing Disorder, Phys. Rev. Lett. **115**, 077001 (2015).
[8] X. Xu, F.-T. Huang, A. S. Admasu, J. Kim, K. Wang, E. Feng, H. Cao, and S.-W. Cheong, Bilayer Square Lattice $Tb_2SrAl_2O_7$ with Structural $Z_8$ Vortices and Magnetic Frustration, Chem. Mater. **34**, 1225 (2022).
[9] J. A. Wilson and A. D. Yoffe, The transition metal dichalcogenides discussion and interpretation of the observed optical, electrical and structural properties, Adv. Phys. **18**, 193 (1969).
[10] H. Luo, W. Xie, J. Tao, H. Inoue, A. Gyenis, J. W. Krizan, A. Yazdani, Y. Zhu, and R. J. Cava, Polytypism, polymorphism, and superconductivity in $TaSe_{2−x}Te_x$, Proceedings of the National Academy of Sciences **112**, E1174 (2015).
[11] T. A. Empante, Y. Zhou, V. Klee, A. E. Nguyen, I. H. Lu, M. D. Valentin, S. A. Naghibi Alvillar, E. Preciado, A. J. Berges, C. S. Merida, M. Gomez, S. Bobek, M. Isarraraz, E. J. Reed, and L. Bartels, Chemical Vapor Deposition Growth of Few-

Layer MoTe2 in the 2H, 1T′, and 1T Phases: Tunable Properties of MoTe2 Films, ACS Nano **11**, 900 (2017).

[12] X. Mi, X. Wang, H. Gui, M. Pi, T. Zheng, K. Yang, Y. Gan, P. Wang, A. Li, A. Wang, L. Zhang, Y. Su, Y. Chai, and M. He, Stacking faults in α-RuCl3 revealed by local electric polarization, Phys. Rev. B **103**, 174413 (2021).

[13] H. Zhou, T. Xie, T. Taniguchi, K. Watanabe, and A. F. Young, Superconductivity in rhombohedral trilayer graphene, Nature **598**, 434 (2021).

[14] E. A. Zvereva, G. V. Raganyan, T. M. Vasilchikova, V. B. Nalbandyan, D. A. Gafurov, E. L. Vavilova, K. V. Zakharov, H. J. Koo, V. Y. Pomjakushin, A. E. Susloparova, A. I. Kurbakov, A. N. Vasiliev, and M. H. Whangbo, Hidden magnetic order in the triangular-lattice magnet Li2MnTeO6, Phys. Rev. B **102**, 094433 (2020).

[15] P. M. Woodward, A. W. Sleight, L.-S. Du, and C. P. Grey, Structural Studies and Order–Disorder Phenomenon in a Series of New Quaternary Tellurates of the Type A2+M4+Te6+O6 and A1+2M4+Te6+O6, J. Solid State Chem. **147**, 99 (1999).

[16] A. I. Kurbakov, A. E. Susloparova, V. Y. Pomjakushin, Y. Skourski, E. L. Vavilova, T. M. Vasilchikova, G. V. Raganyan, and A. N. Vasiliev, Commensurate helicoidal order in the triangular layered magnet Na2MnTeO6, Phys. Rev. B **105**, 064416 (2022).

[17] V. B. Nalbandyan, I. L. Shukaev, G. V. Raganyan, A. Svyazhin, A. N. Vasiliev, and E. A. Zvereva, Preparation, Crystal Chemistry, and Hidden Magnetic Order in the Family of Trigonal Layered Tellurates A2Mn(4+)TeO6 (A = Li, Na, Ag, or Tl), Inorg. Chem. **58**, 5524 (2019).

[18] D. Wang, Y. Zhang, Q. Liu, B. Zhang, D. Yang, and Y. Wang, Band gap modulation and nonlinear optical properties of quaternary tellurates Li2GeTeO6, Dalton Trans. **51**, 8955 (2022).

[19] G. M. Sheldrick, Crystal structure refinement with SHELXL, Acta Crystallographica Section C: Structural Chemistry **71**, 3 (2015).

[20] G. M. Sheldrick, SHELXT–Integrated space-group and crystal-structure determination, Acta Crystallographica Section A: Foundations and Advances **71**, 3 (2015).

[21] See Supplemental Material at [URL will be inserted by publisher] for (00L) peaks in XRD, STEM on stacking fault area, inverse susceptibility, M/H, AC susceptibility, and comparison of heat capacity between 2H and 3R phases.

[22] L. Ortega-San Martin, J. P. Chapman, L. Lezama, J. Sánchez Marcos, J. Rodríguez-Fernández, M. I. Arriortua, and T. Rojo, Magnetic Properties of the Ordered Double Perovskite Sr2MnTeO6, Eur. J. Inorg. Chem. **2006**, 1362 (2006).

[23] L. Li, X. Hu, Z. Liu, J. Yu, B. Cheng, S. Deng, L. He, K. Cao, D.-X. Yao, and M. Wang, Structure and magnetic properties of the S = 3/2 zigzag spin chain antiferromagnet BaCoTe2O7, Science China Physics, Mechanics & Astronomy **64**, 287412 (2021).

[24] R. Zhong, M. Chung, T. Kong, L. T. Nguyen, S. Lei, and R. J. Cava, Field-induced spin-liquid-like state in a magnetic honeycomb lattice, Phys. Rev. B **98**, 220407 (2018).

[25] S. H. Baek, S. H. Do, K. Y. Choi, Y. S. Kwon, A. U. B. Wolter, S. Nishimoto, J.

van den Brink, and B. Büchner, Evidence for a Field-Induced Quantum Spin Liquid in α-RuCl3, Phys. Rev. Lett. **119**, 037201 (2017).

[26] B. Gao, T. Chen, D. W. Tam, C.-L. Huang, K. Sasmal, D. T. Adroja, F. Ye, H. Cao, G. Sala, M. B. Stone, C. Baines, J. A. T. Verezhak, H. Hu, J.-H. Chung, X. Xu, S.-W. Cheong, M. Nallaiyan, S. Spagna, M. B. Maple, A. H. Nevidomskyy, E. Morosan, G. Chen, and P. Dai, Experimental signatures of a three-dimensional quantum spin liquid in effective spin-1/2 Ce2Zr2O7 pyrochlore, Nat. Phys. **15**, 1052 (2019).

[27] R. Gummow, M. Thackeray, W. David, and S. Hull, Structure and electrochemistry of lithium cobalt oxide synthesised at 400 C, Mater. Res. Bull. **27**, 327 (1992).

[28] E. Rossen, J. Reimers, and J. Dahn, Synthesis and electrochemistry of spinel LT LiCoO2, Solid State Ionics **62**, 53 (1993).

[29] H. Wang, Y. I. Jang, B. Huang, D. R. Sadoway, and Y. M. Chiang, TEM study of electrochemical cycling-induced damage and disorder in LiCoO2 cathodes for rechargeable lithium batteries, J. Electrochem. Soc. **146**, 473 (1999).

[30] T. Tsuruhama, T. Hitosugi, H. Oki, Y. Hirose, and T. Hasegawa, Preparation of layered-rhombohedral LiCoO2 epitaxial thin films using pulsed laser deposition, Applied Physics Express **2**, 085502 (2009).

[31] Y. Takahashi, N. Kijima, K. Dokko, M. Nishizawa, I. Uchida, and J. Akimoto, Structure and electron density analysis of electrochemically and chemically delithiated LiCoO2 single crystals, J. Solid State Chem. **180**, 313 (2007).

[32] M.-H. Zhao, W. Wang, Y. Han, X. Xu, Z. Sheng, Y. Wang, M. Wu, C. P. Grams, J. Hemberger, D. Walker, M. Greenblatt, and M.-R. Li, Reversible Structural Transformation between Polar Polymorphs of Li2GeTeO6, Inorg. Chem. **58**, 1599 (2019).

[33] X. Xu, F.-T. Huang, A. S. Admasu, M. Kratochvílová, M.-W. Chu, J.-G. Park, and S.-W. Cheong, Multiple ferroic orders and toroidal magnetoelectricity in the chiral magnet BaCoSiO4, Phys. Rev. B **105**, 184407 (2022).

[34] A. Kitaev, Anyons in an exactly solved model and beyond, Annals of Physics **321**, 2 (2006).

[35] R. Zhong, S. Guo, G. Xu, Z. Xu, and R. J. Cava, Strong quantum fluctuations in a quantum spin liquid candidate with a Co-based triangular lattice, Proceedings of the National Academy of Sciences **116**, 14505 (2019).

[36] R. Zhong, T. Gao, N. P. Ong, and R. J. Cava, Weak-field induced nonmagnetic state in a Co-based honeycomb, Sci. Adv. **6**, eaay6953 (2020).

[37] W. Yao and Y. Li, Ferrimagnetism and anisotropic phase tunability by magnetic fields in Na2Co2TeO6, Phys. Rev. B **101**, 085120 (2020).

[38] G. Lin, J. Jeong, C. Kim, Y. Wang, Q. Huang, T. Masuda, S. Asai, S. Itoh, G. Günther, M. Russina, Z. Lu, J. Sheng, L. Wang, J. Wang, G. Wang, Q. Ren, C. Xi, W. Tong, L. Ling, Z. Liu, L. Wu, J. Mei, Z. Qu, H. Zhou, X. Wang, J.-G. Park, Y. Wan, and J. Ma, Field-induced quantum spin disordered state in spin-1/2 honeycomb magnet Na2Co2TeO6, Nat. Commun. **12**, 5559 (2021).

[39] X. Bai, J. A. M. Paddison, E. Kapit, S. M. Koohpayeh, J. J. Wen, S. E. Dutton, A.

T. Savici, A. I. Kolesnikov, G. E. Granroth, C. L. Broholm, J. T. Chalker, and M. Mourigal, Magnetic Excitations of the Classical Spin Liquid MgCr2O4, Phys. Rev. Lett. **122**, 097201 (2019).

[40] F. A. Garcia, U. F. Kaneko, E. Granado, J. Sichelschmidt, M. Hölzel, J. G. S. Duque, C. A. J. Nunes, R. P. Amaral, P. Marques-Ferreira, and R. Lora-Serrano, Magnetic dimers and trimers in the disordered S = 3/2 spin system BaTi1/2Mn1/2O3, Phys. Rev. B **91**, 224416 (2015).

[41] E. Kermarrec, R. Kumar, G. Bernard, R. Hénaff, P. Mendels, F. Bert, P. L. Paulose, B. K. Hazra, and B. Koteswararao, Classical Spin Liquid State in the S = 5/2 Heisenberg Kagome Antiferromagnet Li9Fe3(P2O7)3(PO4)2, Phys. Rev. Lett. **127**, 157202 (2021).

[42] S. K. Karna, Y. Zhao, R. Sankar, M. Avdeev, P. C. Tseng, C. W. Wang, G. J. Shu, K. Matan, G. Y. Guo, and F. C. Chou, Sodium layer chiral distribution and spin structure of Na2Ni2TeO6 with a Ni honeycomb lattice, Phys. Rev. B **95**, 104408 (2017).

[43] S.-W. Cheong, D. Talbayev, V. Kiryukhin, and A. Saxena, Broken symmetries, non-reciprocity, and multiferroicity, npj Quantum Mater. **3**, 19 (2018).

[44] W. Jin, E. Drueke, S. Li, A. Admasu, R. Owen, M. Day, K. Sun, S.-W. Cheong, and L. Zhao, Observation of a ferro-rotational order coupled with second-order nonlinear optical fields, Nat. Phys. **16**, 42 (2020).

**Table 1. Single Crystal Structure Refinement for 3R-$Na_2MnTeO_6$ at 298(2) K**

| Refined formula | $Na_2MnTeO_6$ |
|---|---|
| F.W. (g/mole) | 313.62 |
| Space group | $R$-3 |
| Z | 3 |
| $a$ (Å) | 5.215(4) |
| $c$ (Å) | 15.94(2) |
| V (Å$^3$) | 375.4(7) |
| extinction coefficient | 0.0028(10) |
| $R_1$ | 0.0360 |
| $\omega R_2$ ($I > 2\delta(I)$) | 0.0702 |
| goodness of fit | 1.147 |

**Table 2. Atomic Coordinates and Equivalent Isotropic Displacement Parameters from the single-crystal refinement of 3R-Na$_2$MnTeO$_6$ at 298(2) K; space group *R-3***

| Atoms | Wyckoff site | x/a | y/b | z/c | Occupation | $U_{eq}$ |
|---|---|---|---|---|---|---|
| Te | 6c | 2/3 | 1/3 | 0.49554(9) | 0.5 | 0.0093(5) |
| Mn | 6c | 2/3 | 1/3 | 0.49554(9) | 0.5 | 0.0093(5) |
| Na | 6c | 1/3 | 2/3 | 0.3608(4) | 1 | 0.0294(16) |
| O | 18f | 0.3814(12) | 0.3884(11) | 0.4310(4) | 1 | 0.0273(15) |

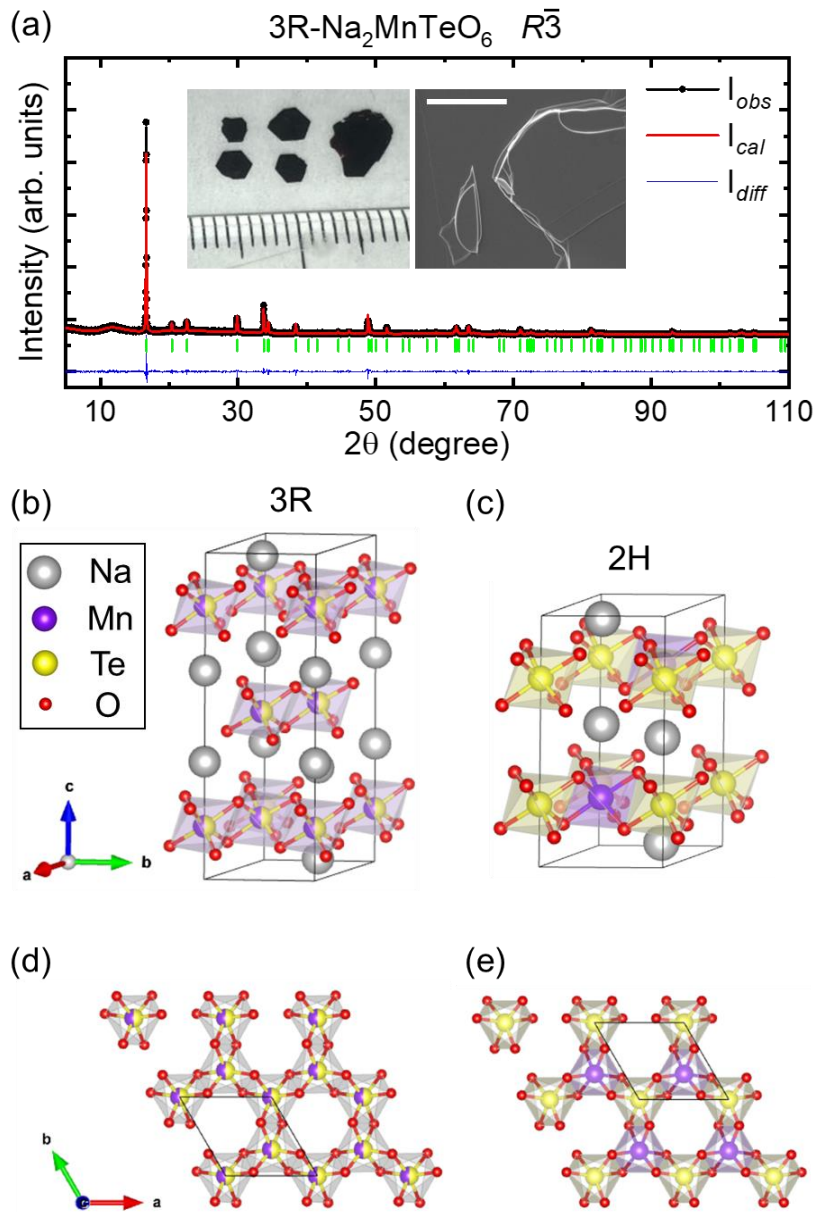

**Figure 1. (a)** The black curve displays the powder XRD collected on the ground powder of 3R-Na$_2$MnTeO$_6$ crystals. The Le-Bail fitting and the difference curves are shown in red and blue, respectively. Green bars exhibit Bragg peak positions. The inset displays a photograph of as-grown 3R-Na$_2$MnTeO$_6$ single crystals and an SEM image taken on the cleaved surface. The white scale bar is 40 μm. **(b)-(c)** The crystal structures of 3R- (b) and 2H- (c) Na$_2$MnTeO$_6$. **(d)-(e)** The schematics of magnetic layers (*ab* planes) of 3R- (d) and 2H- (e) Na$_2$MnTeO$_6$.

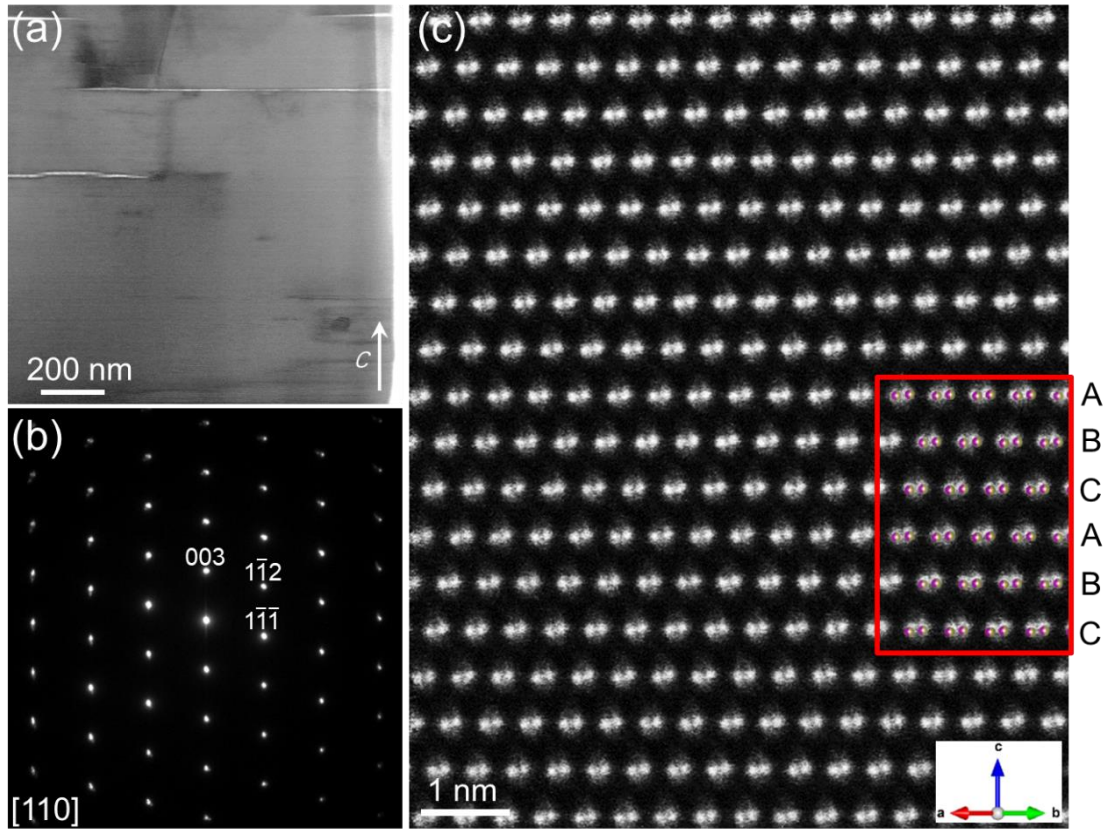

**Figure 2. (a)-(b)** A low-magnification TEM image and corresponding diffraction pattern. **(c)** The STEM atomic image showing the Mn-Te atom arrays. the *R*-3 Mn-Te disordered structural model is superimposed on the region marked by the red square (Mn, purple; Te, yellow). All data are taken with electron beam parallel to the [110] direction.

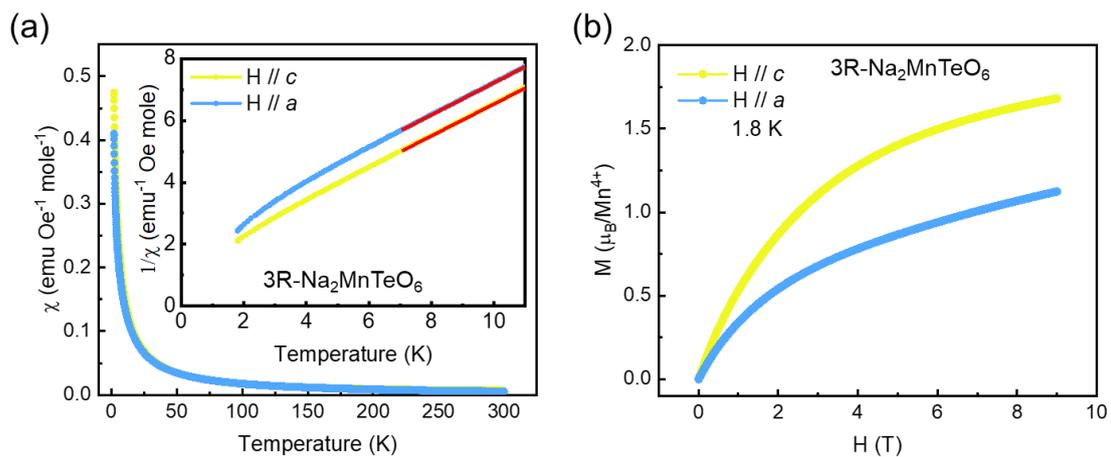

**Figure 3. (a)** The magnetic susceptibility of 3R-Na$_2$MnTeO$_6$ versus temperature curves measured by 1000 Oe field along *c* and *a* are displayed in yellow and blue, respectively. The inset shows the low-temperature range inverse susceptibility and the CW fitting (red curves). See Fig. S3 for the high-temperature range [21]. **(b)** The isothermal magnetization versus magnetic field data at 1.8 K.

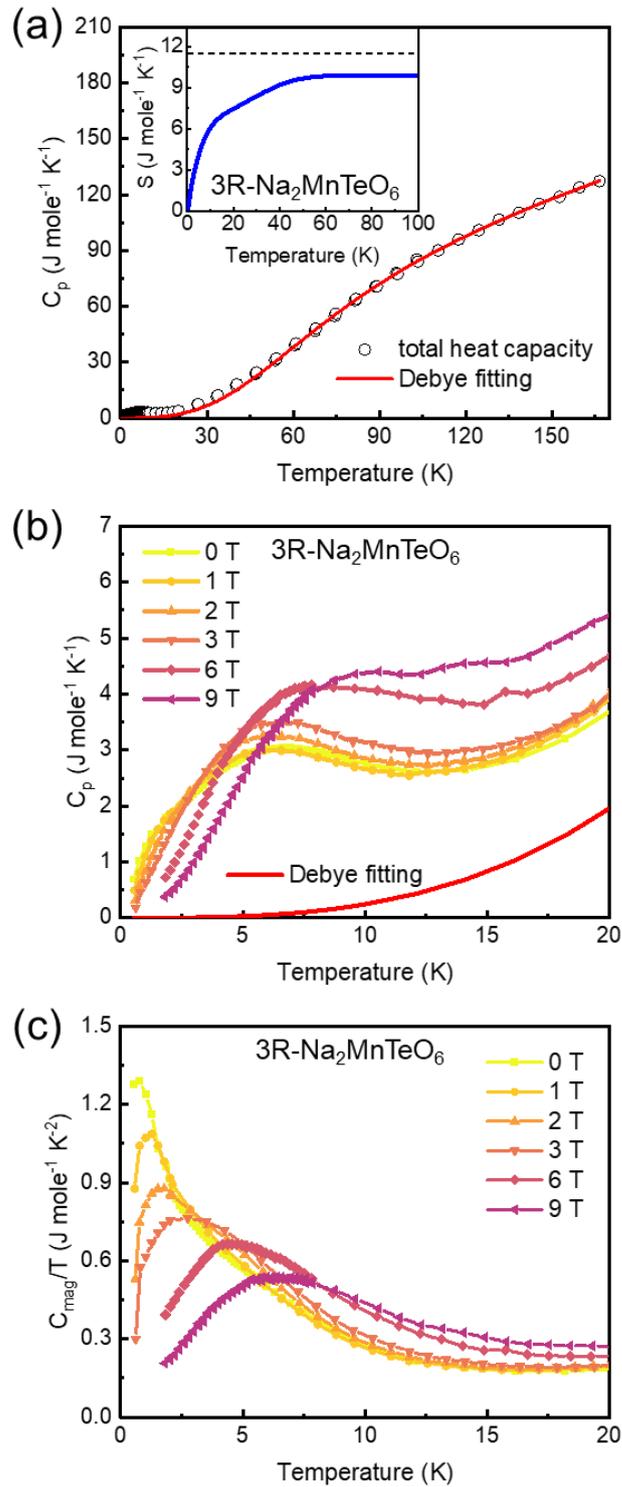

**Figure 4. (a)** Black open circles show the heat capacity of 3R-Na$_2$MnTeO$_6$ as a function of temperature. The red curve indicates the Debye fitting of the high temperature range. The inset displays the magnetic entropy. **(b)-(c)** The heat capacity and magnetic heat capacity over temperature measured at various fields applied along *c* axis.

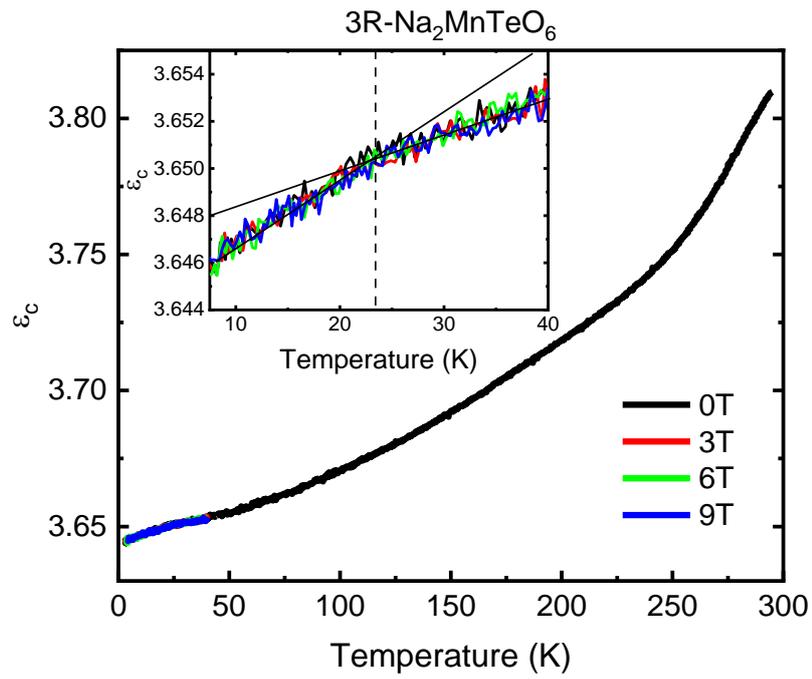

**Figure 5.** The temperature-dependent dielectric permittivity of 3R-$Na_2MnTeO_6$ along *c* axis ($\varepsilon_c$) measured with 44 kHz frequency and 1 V AC voltage, and at 0 T, 3 T, 6 T, 9 T magnetic field along *c* displayed by the black, red, green, and blue curves, respectively.

*Supplementary Information*

**Spin disorder in a stacking polytype of a layered magnet**
Xianghan Xu[1*], Guangming Cheng[2], Danrui Ni[1], Xin Gui[3], Weiwei Xie[4], Nan Yao[2], and R. J. Cava[1*]

1. Department of Chemistry, Princeton University, Princeton, NJ 08544
2. Princeton Institute for the Science and Technology of Materials, Princeton University, Princeton, NJ 08544
3. Department of Chemistry, University of Pittsburgh, Pittsburgh, PA 15260
4. Department of Chemistry, Michigan State University, Lansing, MI 48824

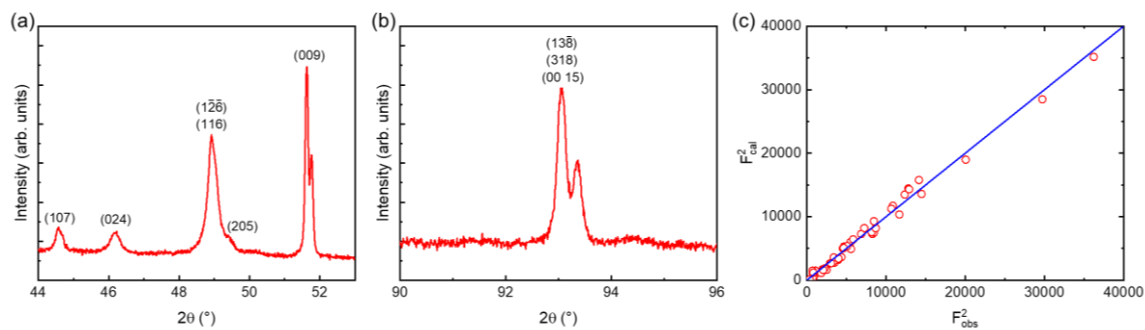

**Figure S1**. **(a)-(b)** The powder XRD pattern on ground 3R-$Na_2MnTeO_6$ in 2θ regimes containing (009) and (00 15) diffraction peaks, respectively. The peak indexes are marked correspondingly in the figure. (c) The calculated intensity square vs. observed intensity square for 3R-$Na_2MnTeO_6$ SC-XRD refinement.

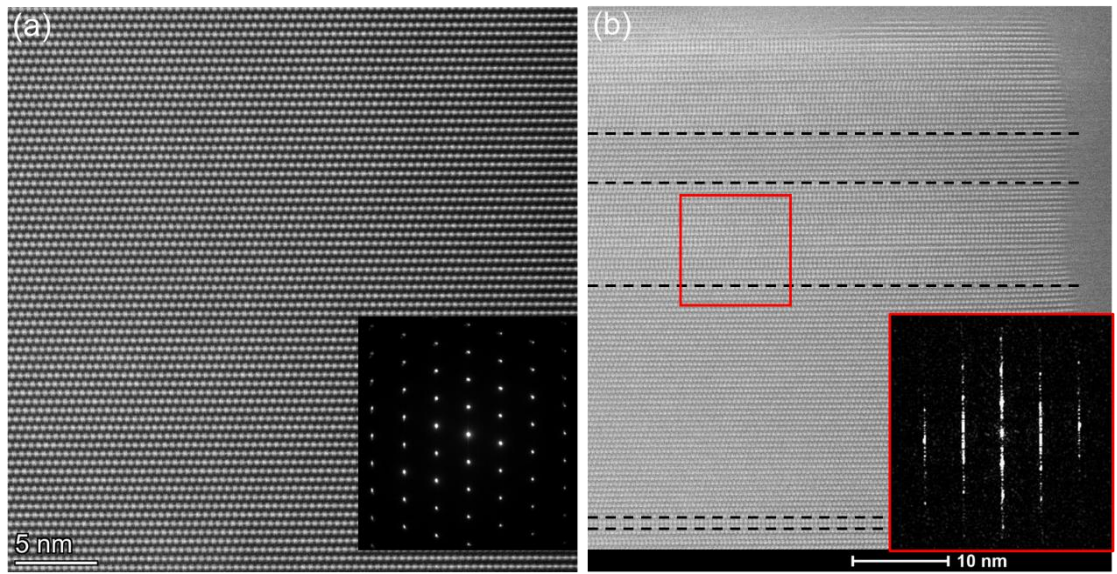

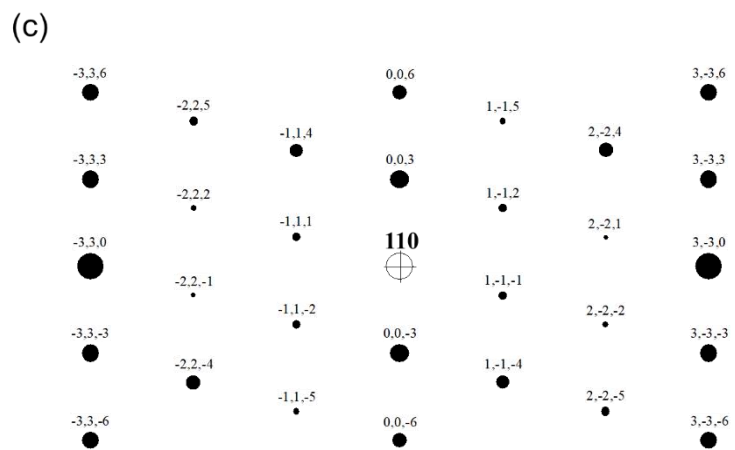

**Figure S2.** The STEM atomic imaging of **(a)** an area with perfect ABC stacking and **(b)** an area with stacking faults. The boundaries between ABC stacking and stacking faults are marked by the black dashed lines. The corresponding diffraction patterns are displayed in the insets. **(c)** The simulated electron diffraction pattern using the *R*-3 structural model.

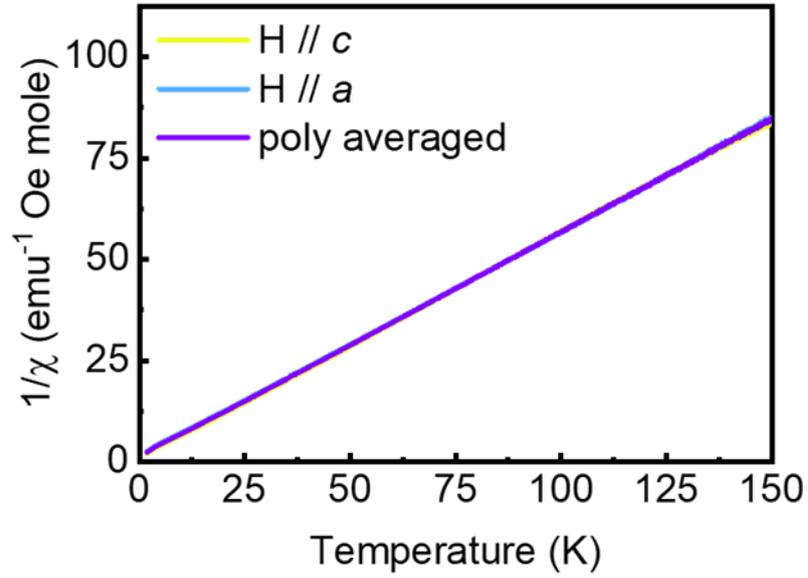

**Figure S3**. The inverse susceptibility as a function of temperature curves for H//*c* (yellow), H//*a* (light blue) on a single crystal, and the resulting calculated average (purple).

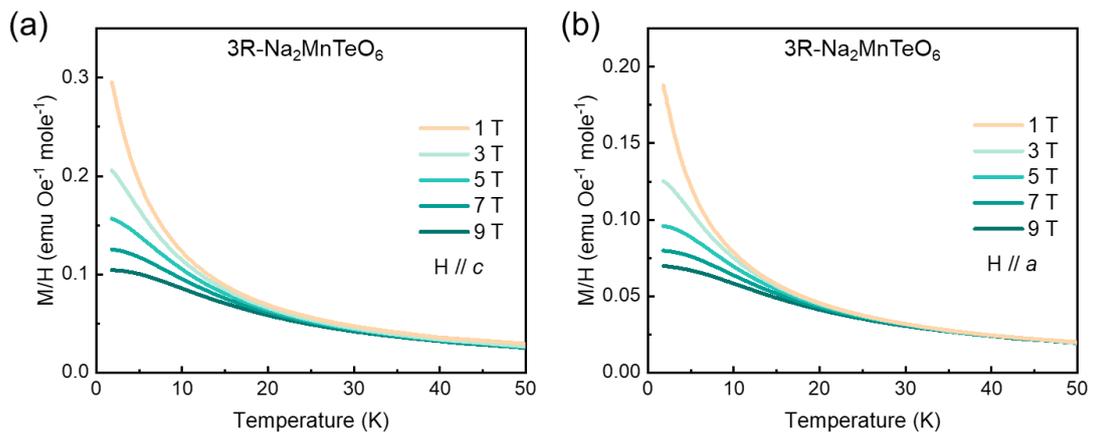

**Figure S4. (a)-(b)** M/H as a function of temperature curves measured with 1T, 3T, 5T, 7T, 9T fields applied along *c* and *a* directions, respectively.

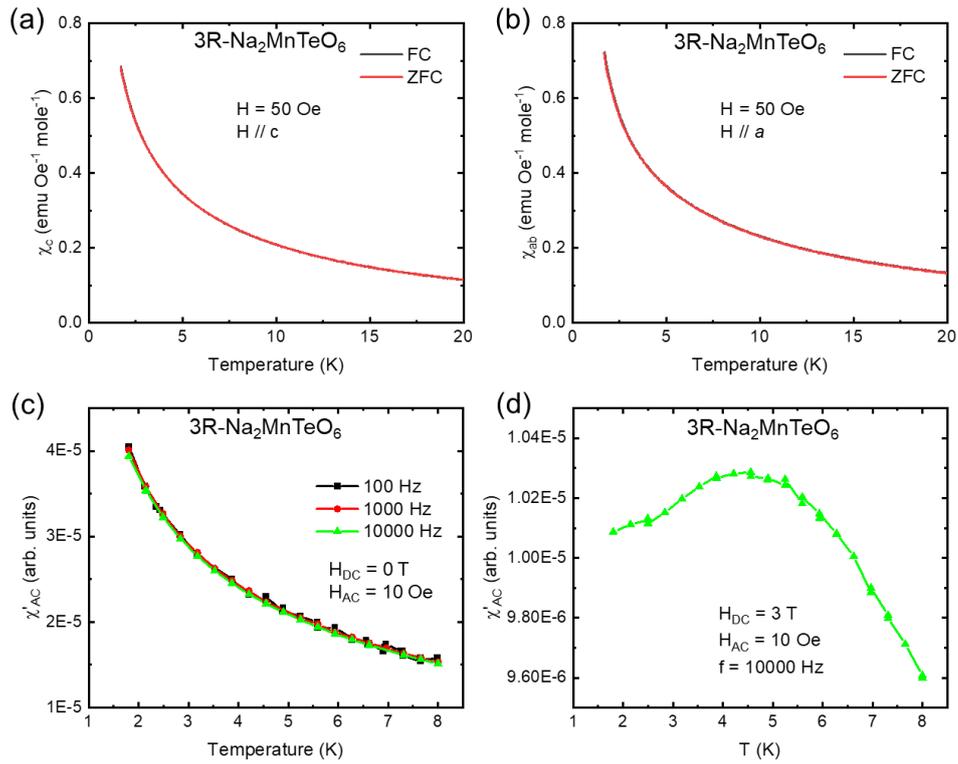

**Figure S5. (a)-(b)** The FC (black) and ZFC (red) magnetic susceptibility as a function of temperature measured in a tiny field (50 Oe) along $c$ and $a$ directions, respectively. **(c)** The AC susceptibility as a function of temperature measured in 100 Hz, 1000 Hz, and 10000 Hz AC fields, and zero DC field along $c$ direction. **(d)** The AC susceptibility as a function of temperature measured in 10,000 Hz AC field and 3T DC field along $c$ direction.

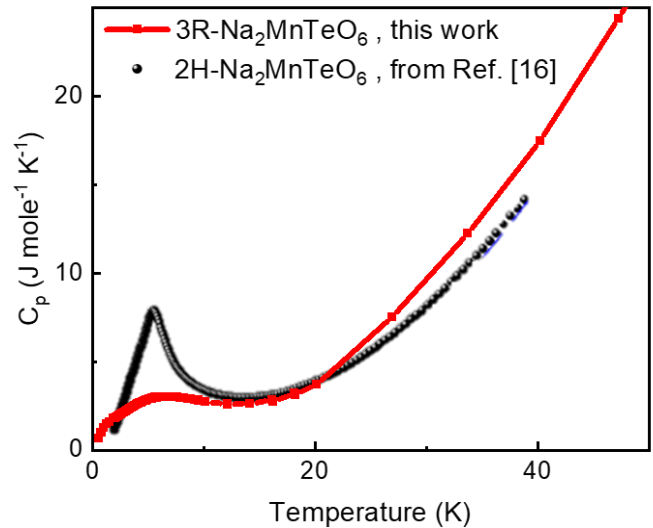

**Figure S6**. The comparison of the heat capacity measured on 3R-$Na_2MnTeO_6$ and 2H-$Na_2MnTeO_6$ (extracted from Ref. [16] of the main text).